\input{aipcheck}

\documentclass[
    ,final            
  ]
  {aipproc}

\layoutstyle{6x9}

\usepackage{overpic}
\usepackage{subfigure}
\usepackage[utf8]{inputenc}
\usepackage[english]{babel}
\usepackage{amsmath}
\usepackage{amsfonts}
\usepackage{amssymb}
\usepackage{epsfig}
\usepackage{graphics,psfrag,rotating}
\usepackage{graphicx}
\usepackage{dcolumn}
\usepackage{bm}

\begin{document}

\title{On the collapse in fourth order gravities}

\classification{
98.80.-k 
, 04.50.-h 
              }
\keywords      {Modified Gravity, cosmology}

\author{B.\,Montes N\'u\~nez}{
address={Centro de Investigaciones Energ\'eticas, Medioambientales y Tecnol\'ogicas (CIEMAT), 28040 Madrid, Spain.} 
}

\author{J.\,A.\,R.\,Cembranos}{
  address={Departamento de F\'{\i}sica Te\'orica I, Universidad Complutense de Madrid, E-28040 Madrid, Spain.}
}

\author{A.\,de la Cruz-Dombriz}{
  address={Astrophysics, Cosmology and Gravity Centre (ACGC), University of Cape Town, Rondebosch, 7701, South Africa.}
 ,altaddress={ Department of Mathematics and Applied Mathematics, University of Cape Town, 7701 Rondebosch, Cape Town, South Africa.}
}

\begin{abstract}
The gravitational collapse in fourth order theories of gravity defined by an arbitrary action of the scalar curvature shows
significant deviations with General Relativity. The presence of a new scalar mode produces a higher initial
contraction that favors the reduction of the collapsing time. However, depending on the particular model, there are
fundamental differences when the modifications to the General Relativity collapse leave the linear regime. These analyses
can be used to exclude an important region of the parameter space associated with alternative gravitational models.
\end{abstract}

\maketitle


\section{Introduction}
The gravitational collapse for a spherically symmetric stellar object has been analyzed in detail in the standard General Relativity (GR) theory \cite{Weinberg}. By assuming the metric of the space-time to be spherically symmetric and that the collapsing fluid is pressureless, the found metric interior to the object turns to be Robertson-Walker type with a parameter playing the role of spatial curvature and proportional to initial density. The time lapse and the size of the object are given by  a cycloid parametric equation with an angle parameter $\psi$.

In spite of the fact that GR has been one of the most successful theories of the twentieth century, it does not give a satisfactory explanation to some of the latest cosmological and astrophysical observations with usual matter sources. On the one hand, the baryonic matter content has to be supplemented by a dark matter (DM) component. Although there are many possible origins for this additional component \cite{DM}, DM is usually assumed to be in the form of thermal relics that naturally freeze-out with the right abundance in many extensions of the standard model of particles \cite{WIMPs}. Future experiments will be able to discriminate among the large number of candidates and models, such as direct and indirect detection designed explicitly for their search \cite{isearches}, or even at high energy colliders, where they could be produced \cite{Coll}.

On the other hand, a dark energy contribution needs to be considered to provide cosmological acceleration \cite{DE}. Alternatively,
a modified gravitational interaction can potentially generate the present expansion of the Universe without invoking the presence of any exotic dark energy \cite{otras}. In this context, fourth order theories of gravity defined by adding to the action a function of the scalar curvature:
\begin{equation}
S_G=\frac{1}{16\pi G}\int \text{d}^4x \sqrt{\vert g \vert}\left(R+f(R)\right)\,,
\label{Modified action}
\end{equation}
has been extensively studied during the last years \cite{R2a,Rm1,varia,R2b}.

\section{Gravitational collapse}
\begin{figure}[tp]
	\centering
		\includegraphics[width=0.48\textwidth]{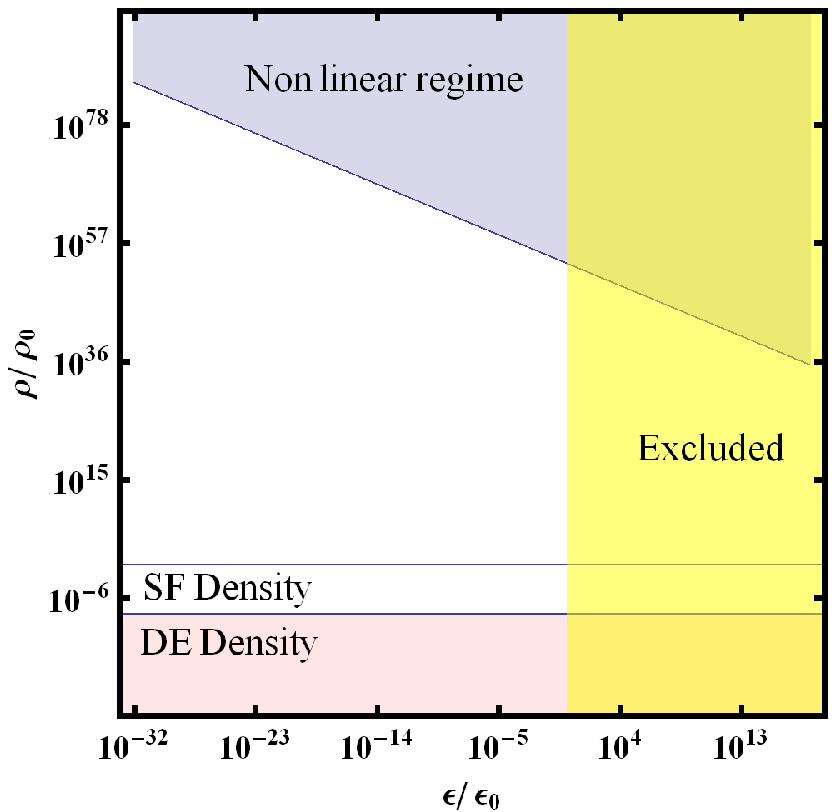}\;\;\;\;
		\includegraphics[width=0.48\textwidth]{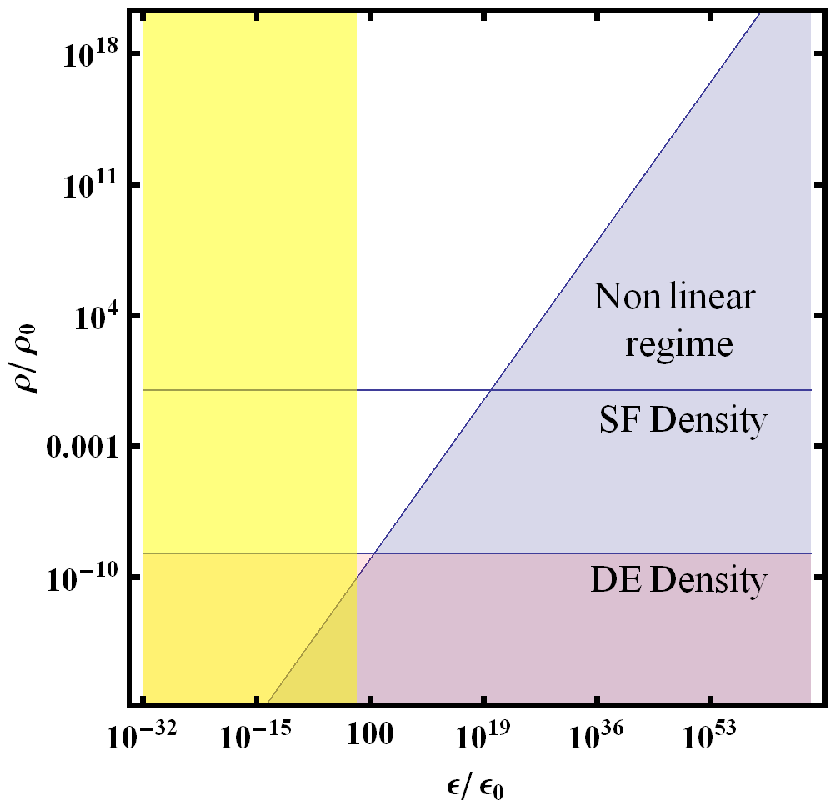}
		\caption{\footnotesize{
Validity of the perturbative regime for $f(R)= \epsilon R^2$ \cite{R2a,R2b} (left panel) and $f(R)= \epsilon R^{-1}$ \cite{Rm1} (right panel). As it is indicated in the figures, in the first case, it is the high density region, the one that exhibits a non linear regime; whereas it is the low density region for the second model (on the right). The right area on the left panel is excluded by the condition $\varepsilon\leq 2.3\times10^{22}\,\text{GeV}^{-2}$ \cite{R2b}. Finally, the density marking the beginning of structure formation (SF) $\rho_{\text{SF}}\simeq1.5 \times 10^{-38}$ GeV$^4$, and the dark energy (DE) density $\rho_{\text{DE}}\simeq2.8 \times 10^{-47}$ GeV$^4$, have also been plotted for reference \cite{Cembranos:2012fd}.}}
\label{validity}
\end{figure}

In order to study the gravitational collapse inside this modified gravity theories, so-called $f(R)$ theories, we can make use of the spherically symmetric metric. The collapsing object is considered to be approximated pressureless and homogeneous. Therefore, we can search a separable solution for this metric as follows \cite{Weinberg}:
\begin{eqnarray}
&&\text{d}s^2 =\text{d}t^2-U(r,t)\text{d}r^2-V(r,t)(\text{d}\theta^2+\text{sin}^2\theta \text{d}\phi^2),\\
&&U(r,t)=A^2(t)h(r),\quad V(r,t)=A^2(t)r^2.
\label{Intervalo colapso, U y V}
\end{eqnarray}
Components $tt$, $rr$ and $\theta\theta$ for the modified tensorial equations may be written respectively in terms of the functions $A(t)$ and $h(r)$. Once we have calculated $h(r)$, the resulting metric is formally the same as the one obtained in GR. We consider $\rho\,(t)=\rho\,(t=0)/A(t)^3$ (given by the energy motion equation for dust matter). Furthermore, provided that the fluid is assumed to be at rest for $t=0$, initial conditions $\dot{A}(t=0)=0$ and $A(t=0)=1$ hold. We define $R(t=0)\equiv R_0$ and $\rho\,(t=0)\equiv \rho_0$ to simplify notation. Taking into account the previous conditions, the evolution equation for $A$ is found to be \cite{Cembranos:2012fd}:
\begin{eqnarray}
\dot{A}^2&=&-\frac{1}{6(1-f_R^2(R_0))}\big[\vphantom{\ddot{f}_R(R_0)}\,8\pi G\rho_0\left(2-f_R(R_0)\right)
- f(R_0)(1+f_R(R_0))
\\
&&
-3\ddot{f}_R(R_0)f_R(R_0)\,\big]
+\frac{1}{(1-f_R^2)}\,\frac{8\pi G}{3}\rho_0\,A^{-1}
-\frac{1}{6(1-f_R^2)}\big[8\pi G\rho_0\,A^{-1}f_R
\nonumber\\
&&
+3A^2\ddot{f}_{R}f_R-3A\dot{A}\dot{f}_{R}(1-f_R)+A^2f(R)(1+f_R)\big]\,.\nonumber
\label{Simplified Eq for dotA}
\end{eqnarray}
Let us remind at this stage that the zeroth order solution of GR is given by the parametric equations of a cycloid \cite{Weinberg}:
\begin{equation}
t=\frac{\psi+\text{sin}\,\psi}{2\sqrt{k}},\;\;\; A_G=\frac{1}{2}(1+\text{cos}\,\psi).
\label{Eqs cicloide}
\end{equation}
Expression \eqref{Eqs cicloide} clearly implies that a sphere with initial density $\rho_0$ and negligible pressure will collapse from rest to a state of infinite proper energy density in a finite time that we will denote $T_G$. This time is obtained for the first value of $\psi$ such as $A_G=0$, i.e. for $\psi=\pi$. In order to study the modification to the gravitational collapse in $f(R)$ theories, we have expanded $A$ around $A_G$ and $f(R)$ around the scalar curvature in GR ($R=R_G$) \cite{Cembranos:2012fd},i.e. $A=A_G+g(\psi)$, and $f(R)\simeq f(R_G)+f'(R_G)(R-R_G)$.
The presence of a function $f(R)$ in the gravitational Lagrangian represents a correction of first order with respect to the usual Einstein-Hilbert action. By substituting the above series expansions in expression \eqref{Simplified Eq for dotA} until first order in $\varepsilon$,  we find that equation (\ref{Simplified Eq for dotA}) becomes \cite{Cembranos:2012fd}:
\begin{eqnarray}
\text{tg}\left(\frac{\psi}{2}\right)g'\,&=&\,-\frac{1}{2}\text{cos}^{-2}\left(\frac{\psi}{2}\right) g
+\frac{1}{12k}\text{cos}^{2}\left(\frac{\psi}{2}\right)\left(f(R_{G0})+3 k f_{R}(R_{G0})\right)
\\
&-&\frac{1}{4}f_R(R_G)-\frac{1}{12k}\text{cos}^{6}\left(\frac{\psi}{2}\right) f(R_G)+\frac{1}{4\sqrt{k}}\,\text{sin}^{}\left(\frac{\psi}{2}\right)\text{cos}^{3}\left(\frac{\psi}{2}\right)\dot{f}_{R}(R_G)
\,. \nonumber
\label{Eq g' with generic Rs}
\end{eqnarray}
By studying this equation for particular $f(R)$ models, it is possible to constraint an important region of the parameter space of
this gravitational theories (See Fig. \ref{validity}). This validity tests are interesting since the higher contraction that these models generally present in the perturvative regime, can alleviate some tension of the standard $\Lambda$CDM model with observations of structures at high redshift \cite{clusters}.

\begin{theacknowledgments}
This work has been supported by MICINN (Spain) project numbers FIS 2008-01323, FIS2011-23000, FPA 2008-00592, FPA2011-27853-01, Consolider-Ingenio MULTIDARK CSD2009-00064 (Spain) and URC (South Africa).
\end{theacknowledgments}

\bibliographystyle{aipproc}   

\begin{thebibliography}{9}

\bibitem{Weinberg}
  S.~Weinberg, Gravitation and Cosmology: Principles and Applications of the General Theory of Relativity, (1972).

\bibitem{DM}
  L.~Covi, J.~E.~Kim and L.~Roszkowski, Phys.\ Rev.\ Lett.\  {\bf 82}, 4180 (1999);
  J.~L.~Feng, A.~Rajaraman and F.~Takayama, Phys.\ Rev.\ D {\bf 68}, 085018 (2003); 
  J.~L.~Feng, A.~Rajaraman and F.~Takayama,  Int.\ J.\ Mod.\ Phys.\ D {\bf 13}, 2355 (2004); 
  J.~A.~R.~Cembranos, J.~L.~Feng, A.~Rajaraman and F.~Takayama, Phys.\ Rev.\ Lett.\  {\bf 95}, 181301 (2005); 
  J.~A.~R. Cembranos, J.~L.~Feng, L~E.~Strigari,  Phys.\ Rev.\  D {\bf 75}, 036004 (2007); 
  J.~A.~R.~Cembranos, J.~H.~Montes de Oca Y., L.~Prado, J.\ Phys.\ Conf.\ Ser.\  {\bf 315}, 012012 (2011); 
  J.~A.~R.~Cembranos, J.~L.~Diaz-Cruz and L.~Prado, Phys.\ Rev.\ D {\bf 84}, 083522 (2011). 

\bibitem{WIMPs}
  H.~Goldberg, Phys.\ Rev.\ Lett.\  {\bf 50}, 1419 (1983); 
  J.~R.~Ellis {\it et al.}, Nucl.\ Phys.\ B {\bf 238}, 453 (1984); 
  K. Griest and M. Kamionkowski, Phys. Rep. \textbf{333}, 167 (2000);
  J.~A.~R.~Cembranos, A.~Dobado and A.~L.~Maroto,  Phys.\ Rev.\ Lett.\  {\bf 90}, 241301 (2003); 
  Phys.\ Rev.\ D {\bf 68}, 103505 (2003); 
  AIP Conf. Proc. {\bf 670}, 235 (2003); 
  Phys.\ Rev.\ D {\bf 73}, 035008 (2006); 
  Phys.\ Rev.\ D {\bf 73}, 057303 (2006); 
  Int. J. Mod. Phys. {\bf D13}, 2275 (2004); 
  A.~L.~Maroto, Phys.\ Rev.\ D {\bf 69}, 043509 (2004); 
  Phys.\ Rev.\ D {\bf 69}, 101304 (2004); 
  A. Dobado and A. L. Maroto, Nucl. Phys. B \textbf{592}, 203 (2001);
  J.~A.~R.~Cembranos {\it et al.}, JCAP {\bf 0810}, 039 (2008). 

\bibitem{isearches}
  J.~A.~R.~Cembranos and L.~E.~Strigari, Phys.\ Rev.\  D {\bf 77}, 123519 (2008); 
  J.~A.~R.~Cembranos, J.~L.~Feng and L.~E.~Strigari, Phys.\ Rev.\ Lett.\  {\bf 99}, 191301 (2007); 
  J.~A.~R.~Cembranos {\it et al.}, Phys.\ Rev.\  D {\bf 83}, 083507 (2011); 
  J.\ Phys.\ Conf.\ Ser.\  {\bf 314}, 012063 (2011); 
  AIP Conf.\ Proc.\  {\bf 1343}, 595 (2011);  
  arXiv:1111.4448 [astro-ph.CO]; 
  arXiv:1202.1707 [astro-ph.CO]. 

\bibitem{Coll}
  J.~Alcaraz {\it et al.}, Phys. Rev.{\bf D67}, 075010 (2003); 
  P. Achard {\it et al.}, Phys. Lett. {\bf B597}, 145 (2004); 
  J.~A.~R.~Cembranos, A.~Rajaraman and F.~Takayama, Europhys.\ Lett.\  {\bf 82}, 21001 (2008);
  J.~A.~R.~Cembranos, A.~Dobado and A.~L.~Maroto, Phys. Rev. {\bf D65} 026005 (2002); 
  J.\ Phys.\ A  {\bf 40}, 6631 (2007); 
  Phys. Rev. {\bf D70}, 096001 (2004); 
  J.~A.~R.~Cembranos {\it et al.}, AIP Conf.\ Proc.\  {\bf 903}, 591 (2007). 

\bibitem{DE}
  S.~Weinberg, Rev.\ Mod.\ Phys.,\ {\bf 61}, 1-23, (1989);
  T.~Biswas {\it et al.}, Phys.\ Rev.\ Lett.\  {\bf 104}, 021601 (2010); 
  JHEP {\bf 1010}, 048 (2010); 
  Phys.\ Rev.\  D {\bf 82}, 085028 (2010); 
  J.~A.~R.~Cembranos, AIP Conf.\ Proc.\  {\bf 1182}, 288 (2009); 
  Phys.\ Rev.\  D {\bf 73}, 064029 (2006); 
  AIP Conf.\ Proc.\  {\bf 1343}, 604 (2011); 
  J.~A.~R.~Cembranos, K.~A.~Olive, M.~Peloso and J.~P.~Uzan, JCAP {\bf 0907}, 025 (2009); 
  S. Nojiri and S. D. Odintsov, Int. J. Geom. Meth. Mod. Phys. {\bf 4} 115, (2007);
  J. Beltr\'an and A. L. Maroto, Phys. Rev. D {\bf 78}, 063005 (2008);
  JCAP 0903, 016 (2009);
  Phys. Rev. D {\bf 80}, 063512 (2009);
  Int. J. Mod. Phys. D {\bf 18}, 2243-2248 (2009).

\bibitem{otras}
  A. Dobado and A. L. Maroto Phys.\ Rev.\ {\bf D52}, 1895, (1995);
  G. Dvali, G. Gabadadze and M. Porrati, {\it Phys. Lett.} {\bf B485}, 208, (2000);
  S. M. Carroll et al., {\it Phys. Rev.} {\bf D71} 063513, (2005);
  J.~A.~R.~Cembranos, Phys.\ Rev.\  {\bf D73} 064029, (2006);
  S.~Nojiri and S.~D.~Odintsov,  Int.\ J.\ Geom.\ Meth.\ Mod.\ Phys.\  {\bf 4} 115, (2007);   

\bibitem{R2a}
  A.~A.~Starobinsky, Phys.\ Lett.\  B {\bf 91} 99, (1980).

\bibitem{Rm1}
  S.~M.~Carroll, V.~Duvvuri, M.~Trodden, M.~S.~Turner, Phys.\ Rev.\ D {\bf 70 } 043528, (2004). 

\bibitem{varia}
  T.~P.~Sotiriou, Gen.\ Rel.\ Grav.\  {\bf 38} 1407, (2006);  
  O.~Mena, J.~Santiago and J.~Weller, Phys.\ Rev.\ Lett.\  {\bf 96} 041103, (2006);   
  V.~Faraoni, Phys.\ Rev.\   {\bf D74} 023529, (2006);  
  S.~Nojiri and S.~D.~Odintsov, Phys.\ Rev.\  {\bf D74} 086005, (2006);  
  A.~de la Cruz-Dombriz, A.~Dobado, Phys.\ Rev.\  {\bf D74} 087501, (2006); 
  I.~Sawicki and W.~Hu, Phys.\ Rev.\  {\bf D75} 127502, (2007);  
  A.~de la Cruz-Dombriz, A.~Dobado and A.~L.~Maroto,  Phys.\ Rev.\ Lett.\  {\bf 103} 179001 (2009); 
  Phys.\ Rev.\  D {\bf 80} 124011, (2009) [Erratum-ibid.\  D {\bf 83} 029903, (2011)]; 
  J.\ Phys.\ Conf.\ Ser.\  {\bf 229}, 012033 (2010);  
  A.~de la Cruz Dombriz, {\it Some cosmological and astrophysical aspects of modified gravity theories},
  PhD. thesis (2010), [arXiv:1004.5052 [gr-qc]]. ISBN 978-84-693-7628-7;
  J.~A.~R.~Cembranos, A.~de la Cruz-Dombriz and P.~J.~Romero, arXiv:1109.4519 [gr-qc]; 
  arXiv:1202.0853 [gr-qc]; 
  A.~de la Cruz-Dombriz and D.~Saez-Gomez,  arXiv:1112.4481 [gr-qc];
  A.~de la Cruz-Dombriz, A.~Dobado, A.~L.~Maroto,  Phys.\ Rev.\  D{\bf 77 } 123515, (2008); 
  A.~Abebe {\it et al.},  arXiv:1110.1191 [gr-qc];
  S.~Carloni, P.~K.~S.~Dunsby and A.~Troisi, Phys.\ Rev.\ D {\bf 77} 024024 (2008); 
  N.~Goheer, J.~Larena and P.~K.~S.~Dunsby,   Phys.\ Rev.\ D {\bf 80}, 061301 (2009); 
  P.~K.~S.~Dunsby, E.~Elizalde, R.~Goswami, S.~Odintsov and D.~S.~Gomez,   Phys.\ Rev.\ D {\bf 82}, 023519 (2010); 
  A.~M.~Nzioki, S.~Carloni, R.~Goswami and P.~K.~S.~Dunsby,  Phys.\ Rev.\ D {\bf 81}, 084028 (2010). 

\bibitem{R2b}
  J.~A.~R.~Cembranos, Phys.\ Rev.\ Lett.\  {\bf 102}, 141301 (2009); 
  J.\ Phys.\ Conf.\ Ser.\  {\bf 315}, 012004 (2011). 

\bibitem{Cembranos:2012fd}
  J.~A.~R.~Cembranos, A.~de la Cruz-Dombriz and B.~M.~Nunez, arXiv:1201.1289 [gr-qc]. 

\bibitem{clusters}
  R. Foley {\it et al.}, ApJ, 731, 86, (2011);
  M. Brodwin {\it et al.}, ApJ, 721, 90, (2010);
  M. Jee {\it et al.}, ApJ, 704, 672, (2009);
  M. Baldi, V. Pettorino, MNRAS, 412, L1, (2011);
  M. J. Mortonson, W. Hu, D. Huterer, Phys. Rev. D {\bf 83}, 023015, (2011).






\end{thebibliography}

\end{document}